\documentstyle[floats,prl,aps]{revtex} 
 
\input epsf

\begin{document}

\def\degrees{\hbox{${}^\circ$\hskip-3pt .}}
\def\spose#1{\hbox to 0pt{#1\hss}}
\def\simlt{\mathrel{\spose{\lower 3pt\hbox{$\mathchar"218$}}
        \raise 2.0pt\hbox{$\mathchar"13C$}}}
\def\simgt{\mathrel{\spose{\lower 3pt\hbox{$\mathchar"218$}}
     \raise 2.0pt\hbox{$\mathchar"13E$}}}

\newcommand{\Phigb}{\Phi_{\gamma b}}
\newcommand{\RF}{{\cal{T}}}
\newcommand{\Aa}{{\cal A}_a}
\newcommand{\Ab}{{\cal A}_b}
\newcommand{\bg}{{b\gamma}}
\newcommand{\eal}{\!\!\! & = & \!\!\!}

\title{MEASURING THE CURVATURE OF THE UNIVERSE$^{\rm\dag}$}
\author{Wayne Hu${}^1$ \& Martin White${}^2$}
\address{$^{1}$Institute for Advanced
Study, Princeton NJ 08540\\ 
$^{2}$Enrico Fermi Institute, 5640 S.~Ellis Ave,
Chicago IL 60637}

\twocolumn[
\maketitle
\widetext

\begin{abstract}{\baselineskip 0.4cm
We discuss how the curvature of the universe can be robustly
measured employing only the
gross features of the CMB anisotropy spectrum.  
Though the position of the
first peak is not robust, uncertainties in the model for structure
formation can be removed by using the spacing of the acoustic peaks
and the location of the damping tail.  Combined these provide important
consistency tests that can be used to discriminate against a truly
exotic model. 
}
\end{abstract}
\vskip 1.0truecm
]
\narrowtext


If we knew the model of structure formation exactly, we could
determine many cosmological parameters, including the curvature of the
universe, to several percent accuracy 
from features in the CMB anisotropy power spectrum 
(for a review see \cite{naturereview} and references therein). 
The question arises: how is our ability to measure them degraded as we relax
our assumptions about the underlying model.
Once we understand which features are model independent (and why), 
we can go on to study the process of structure formation 
from those which are model {\it dependent}. 
For concreteness, we will focus here on one step of this
program \cite{BigPaper}: measuring the spatial curvature of
the universe, i.e.~$\Omega_{\rm tot} = \Omega_0 + \Omega_\Lambda$.
The complementary approach of first verifying the model and then measuring
the cosmological parameters is taken in a companion piece \cite{Companion}.
\footnotetext{$^{\rm\dag}$To appear in Proceedings of the 
XXXIth Moriond Meeting, {\it Microwave Background Anisotropies}. 
IASSNS-AST-96/38}

Let us assume that we understand the ``big picture'', i.e.~that gravitational
instability enhances initially small fluctuations, 
and that the CMB is coupled
to the baryon-electron plasma before recombination.
Can we build a measurement of the curvature from such minimal assumptions?
Are there sufficient cross checks such that we can have confidence
in the measurement? The answers to these questions lie in the
acoustic signature of the small angle CMB anisotropy spectrum.

Our ``big picture'' leaves several questions unanswered:
\begin{enumerate}
\item
What is the fundamental nature of the fluctuations?
\begin{itemize}
\item[--] do curvature perturbations exist outside the horizon as in the
inflationary model or are the perturbations initially isocurvature as
in a defect model (see also \cite{BigPaper,Companion}).
\end{itemize}
\item What is the matter content of the universe?
\begin{itemize}
\item[--] does the baryon-photon ratio ($\Omega_b h^2$) follow the big
bang nucleosynthesis (BBN) prediction?
\item[--] what is the matter-radiation ratio ($\Omega_0 h^2$ or more generally,
the equality epoch $z_{eq}$)?
\end{itemize}
\item Does the thermal history of the universe follow the prediction of
standard recombination at $z_* \sim 10^3$?  
\end{enumerate}

\noindent
These questions and their consequence for the curvature
measurement can be addressed by examining the 
gross properties of the CMB spectrum taken as
a whole.  The fine details, so useful for making precision measurements
in a fixed model, are too model-dependent to serve us here.

Under our minimal assumptions, we have two striking features
(1) the acoustic peaks: their positions, position ratios, spacings and
relative heights; and (2) the damping tail: its position, position 
relative to the peaks and shape.  The acoustic peaks probe 
the sound horizon at last scattering; the damping tail probes
the photon diffusion scale  
at that epoch.  Both reflect the curvature of the universe in
the projection from physical scale at last scattering to angular scale on
the sky.
We shall show that these features have complementary strengths and
weaknesses in guarding against model uncertainties.  Combined
they can be proof against any one of a host of exotic possibilities.

\begin{figure*}[t]
\begin{center}
\leavevmode
\epsfxsize=3.25in \epsfbox{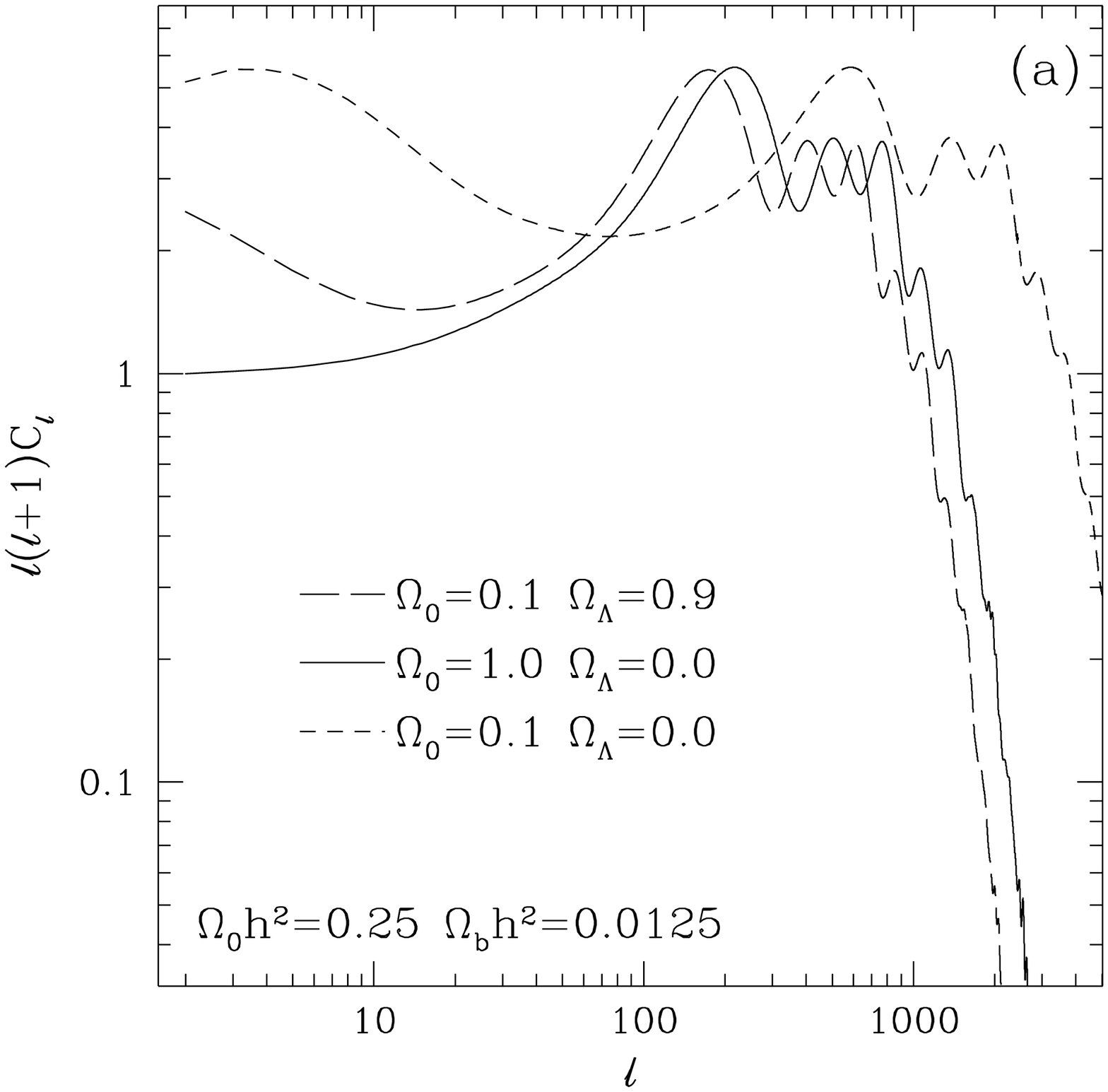}
\epsfxsize=3.25in \epsfbox{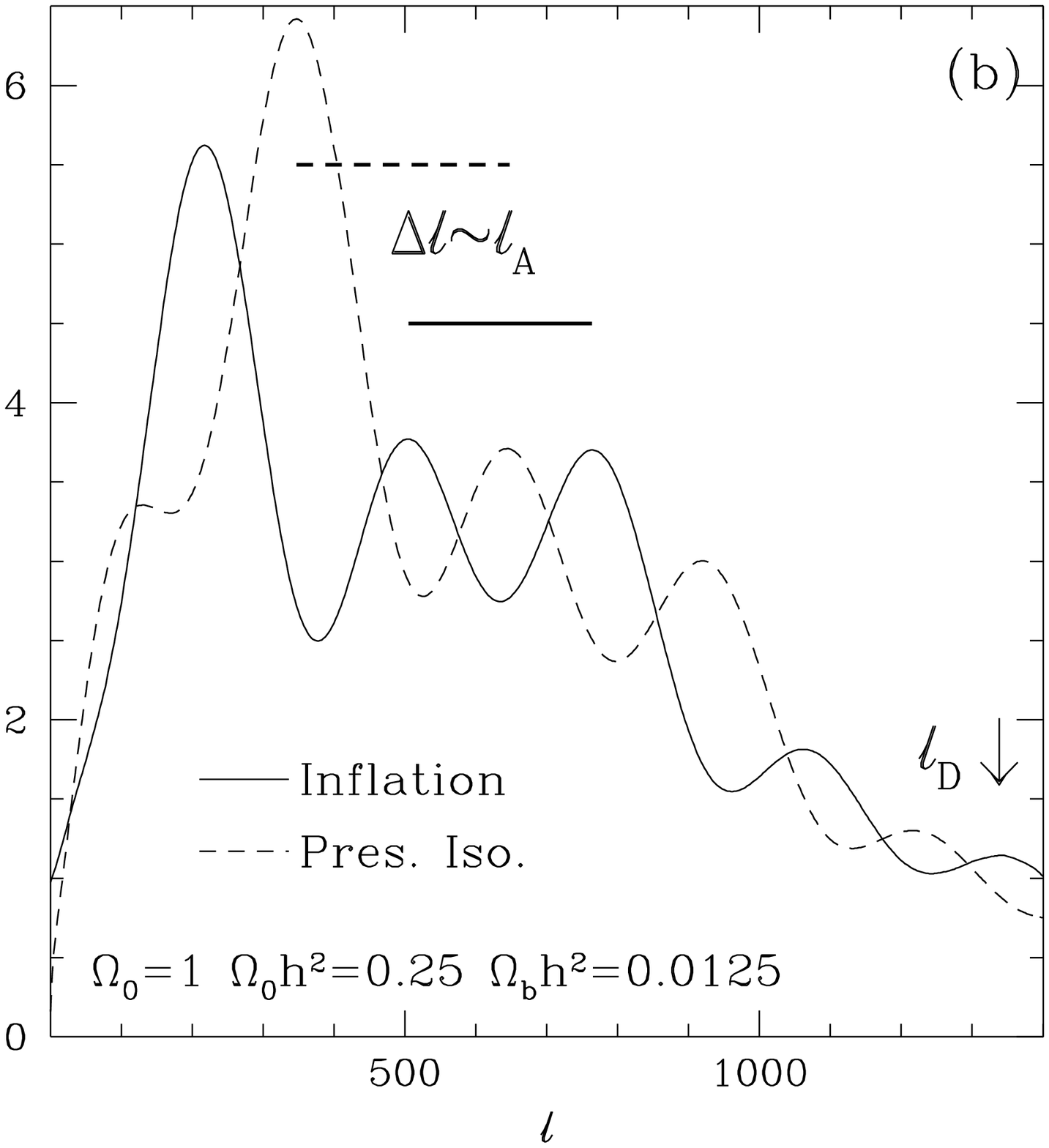}
\end{center}
\caption{\noindent (a) Angular size distance.  For an inflationary
model, features at last scattering such as the peaks and damping
tail are fixed in real space by $\Omega_0 h^2$ and $\Omega_b h^2$, 
providing 
in the anisotropy power spectrum $\ell(\ell+1)C_\ell$ 
standard rulers for the angular size distance test. 
The cosmological constant $\Omega_\Lambda$ yields a minor 
effect compared with the curvature $1-\Omega_0-\Omega_\Lambda$. 
(b) In a broad class of models, 
the peak spacing $\Delta \ell$ and damping 
tail location $\ell_D$ 
depend only on the background parameters and provide
rulers that are robust to model changes. The inflationary
model is here compared with the pressure scaling model 
[7].
}
\label{fig:angle}
\end{figure*}

To set the stage for this discussion, let us briefly review the
angular size distance test for curvature in the universe as it
relates to the CMB spectrum.  A feature in the temperature fluctuations 
on the last scattering surface, 
corresponding to wavenumber $k_{\rm feature}$,
is viewed as an anisotropy on the sky at the multipole moment
of a spherical harmonic decomposition 
$\ell_{\rm feature} = k_{\rm feature} D$, where $D$ is 
the comoving angular size distance and is dependent strongly on
the curvature of the universe:
\begin{equation}
D = |K|^{-1/2} \sinh[|K|^{1/2}(\eta_0-\eta_*)],
\end{equation}
where $\eta_0-\eta_*$ is the conformal distance to the last
scattering surface and the curvature
$K = -H_0^2(1-\Omega_{\rm tot})$ (for $K>0$ replace
$\sinh\rightarrow\sin$).  Here the Hubble constant is
$H_0 = 100 h$ km s$^{-1}$ Mpc$^{-1}$.  Since the projection depends
sensitively on $\Omega_K = 1-\Omega_{\rm tot}$, any feature in
the CMB at last scattering may serve in the angular size distance test
for curvature in the universe (see Fig.~\ref{fig:angle}).  
Let us now turn to the two acoustic features, the peak spacings
and damping tail location, that best suit this purpose.

Before recombination, the photons and baryons are tightly coupled
into a single fluid by Compton scattering.
Acoustic oscillations are stimulated as the gravitational compression
or rarefaction of the fluid is halted and turned around by photon
pressure as the Jeans length (or sound horizon) passes the wavelength.
Because gravity is impotent under the Jeans length, 
typically its effects subsequently die away leaving
the fluid to oscillate at its natural frequency thereafter. 

\begin{figure}
\begin{center}
\leavevmode
\epsfxsize=3.25in \epsfbox{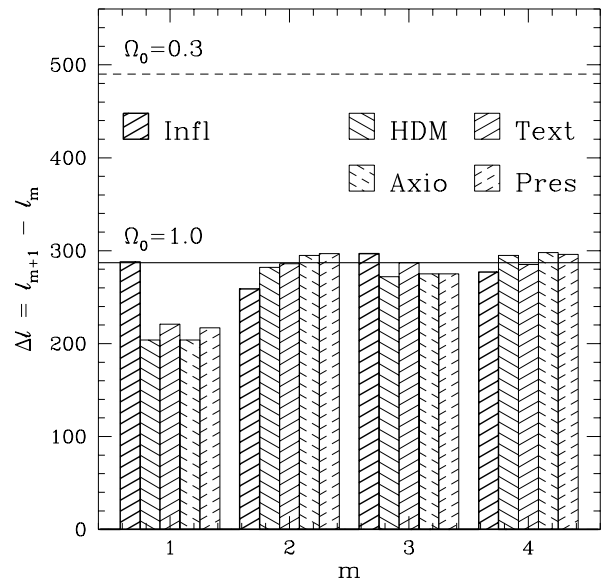}
\end{center}
\caption{\noindent The peak spacing,
 especially between higher peaks, is mainly dependent on the sound horizon 
at last scattering
(projected on the sky) yielding  a robust
feature for the angular size distance test.  
Here five $\Omega_0=1$ models (see text; 
$\Omega_0 h^2 = 0.25$, $\Omega_b h^2=0.0125$ and $\Omega_\Lambda=0$)
are compared with the simple prediction $\Delta \ell = \ell_A$
(solid line).
}
\label{fig:spacing}
\end{figure}

\begin{figure*}
\begin{center}
\leavevmode
\epsfxsize=3.25in \epsfbox{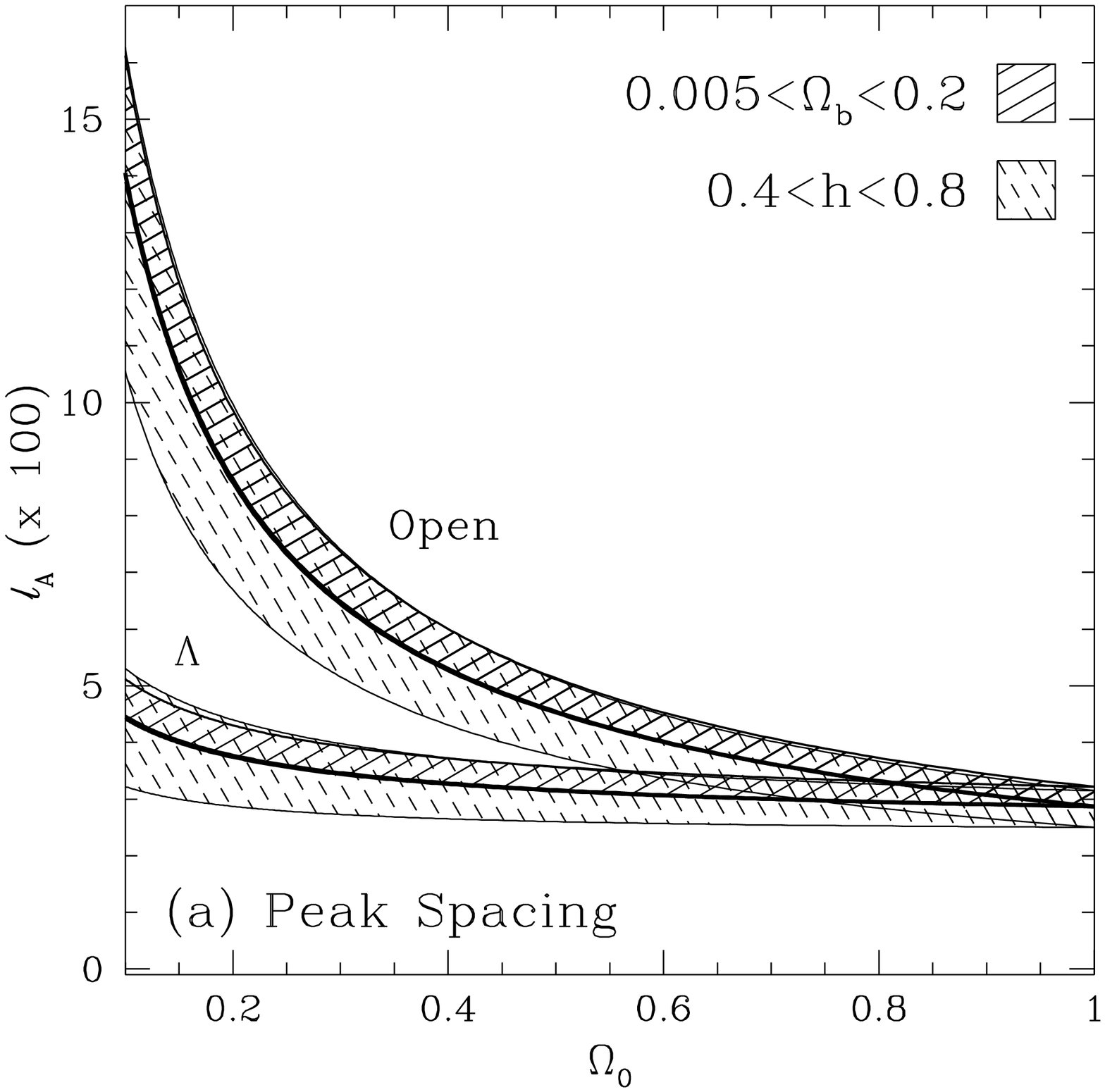} \hskip 0.5truecm
\epsfxsize=3.25in \epsfbox{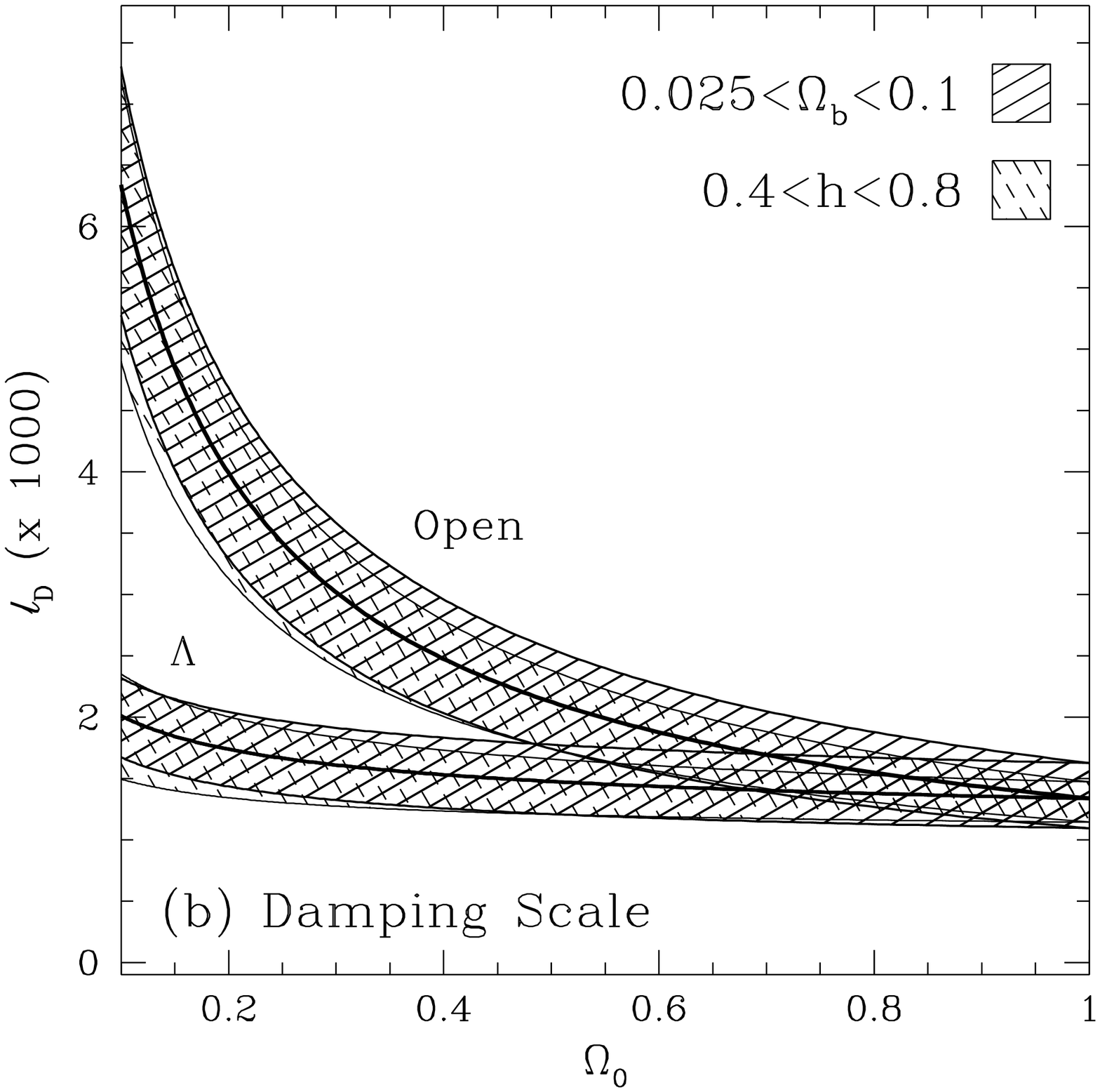}
\end{center}
\caption{\noindent (a) Peak spacing as a function of $\Omega_0$
in a flat cosmological constant ($\Omega_K=0$) 
and open model ($\Omega_\Lambda=0$).
Uncertainties in $\Omega_b h^2$ and $h$ have little effect on the
ability to distinguish an open from a flat model. (b) 
Damping tail as a function of $\Omega_0$.  The damping tail is 
more dependent on $\Omega_b h^2$ than the peak spacing but still
provides interesting constraints.}
\label{fig:omega}
\end{figure*}

As discussed further in \cite{BigPaper,Companion}, 
the nature of the fluctuations 
basically determines whether the photon-baryon fluid is undergoing
a compression or rarefaction at Jeans crossing which affects the
acoustic phase (see Fig.~\ref{fig:angle}b).  
For either case, the spacing of the peaks reflects
the natural frequency of the oscillator.  More specifically,
in the radiation-dominated era the oscillator equation for the
effective temperature fluctuation $T$ of the CMB becomes
$T'' + c_s^2 T = 2\Psi''$,
where $\Psi$ is the Newtonian potential, $c_s$ is the sound
speed, and 
primes are derivatives with respect to $k\eta$, where
$\eta = \int dt/a$ is the conformal time.  From the Poisson equation,
$|\Psi|\sim (k\eta)^{-2} \delta \rho/\rho$. 
Since in the radiation-dominated era, the density fluctuation
$\delta\rho/\rho = {\cal O}(T)$ typically, $\Psi$ is usually
negligible well inside the horizon.
More generally, if $\Psi''$ is small or slowly-varying then 
the solution is an oscillation
at the frequency $\omega = k c_s$, possibly with a zero point
offset.   On the last scattering surface, the acoustic peaks will
be spaced by 
\begin{equation}
k_{m+1}-k_m = k_A = \pi/r_s, \qquad \Delta\ell = \ell_A = k_A D,
\label{eqn:spacing}
\end{equation}
where $r_s = \int c_s d\eta$ 
is the sound horizon at last scattering. 
To summarize: while the phase and first peak bear the mark of the
model-dependent driving force, the spacing of the higher peaks
reflects the model-independent natural frequency, set by the sound
horizon at last scattering.

If the sound horizon at last scattering is a known quantity, then
the peak spacing provides a sensitive angular size distance test
of the curvature that is relatively robust to the nature of the
fluctuations.  In Fig.~\ref{fig:spacing}, we show that the peak spacings 
for the inflationary, texture \cite{CriTur},
hot dark matter \cite{deLSch}, axionic \cite{KawSugYan},
and pressure scaling \cite{HuSpeWhi} isocurvature models are
to good approximation related by Eq.~(\ref{eqn:spacing}) 
to the sound horizon scale.

There are two possible drawbacks to this method of measuring the
curvature.  The first is that the sound horizon at last scattering
depends on the baryon content $\Omega_b h^2$, the matter-radiation
ratio $\Omega_0 h^2$ and the thermal history.  In Fig.~\ref{fig:omega}a,
we show that an uncertain baryon content does not pose an obstacle
nor do reasonable values of the
Hubble constant $0.4 < h < 0.8$.  We shall return to comment on the
thermal history below.  The second drawback is that for a precise
measurement, gravitational forcing effects must be negligible so
that the peak spacing reflects the natural frequency of the oscillator. 
In some models, this may not occur until the higher peaks where
damping and secondary effects may make the signal difficult to observe 
(compare the 1st-2nd peak spacing with the higher ones in
Fig.~\ref{fig:spacing}).

The location of the damping tail in the CMB spectrum
provides yet another angular size distance test of the curvature 
\cite{BigPaper,HuWhi}.  
The damping is a function of the duration of recombination, or the
thickness of the last scattering surface.  As the universe recombines,
the coupling between the photons and the baryons decreases and the
distance that photons can travel before scattering increases.
Acoustic oscillations are destroyed as the photons random walk through
the electron-baryon fluid.
The random walk distance, approximately the geometric mean of the horizon
and Compton mean free path, sets the scale of this feature in the CMB.  

\begin{figure*}
\begin{center}
\leavevmode
\hskip -0.5truecm
\epsfxsize=3.25in \epsfbox{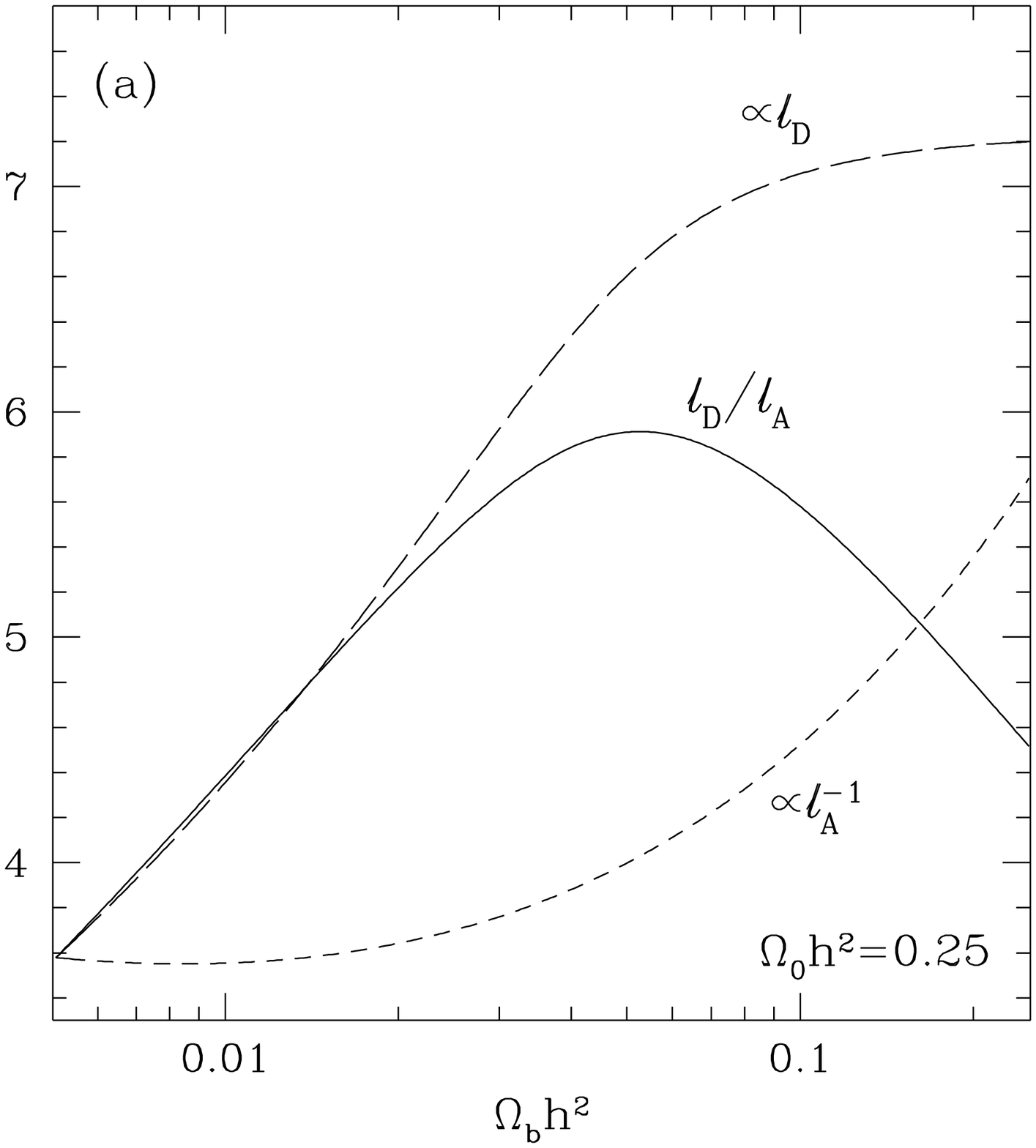} \hskip 0.5truecm
\epsfxsize=3.25in \epsfbox{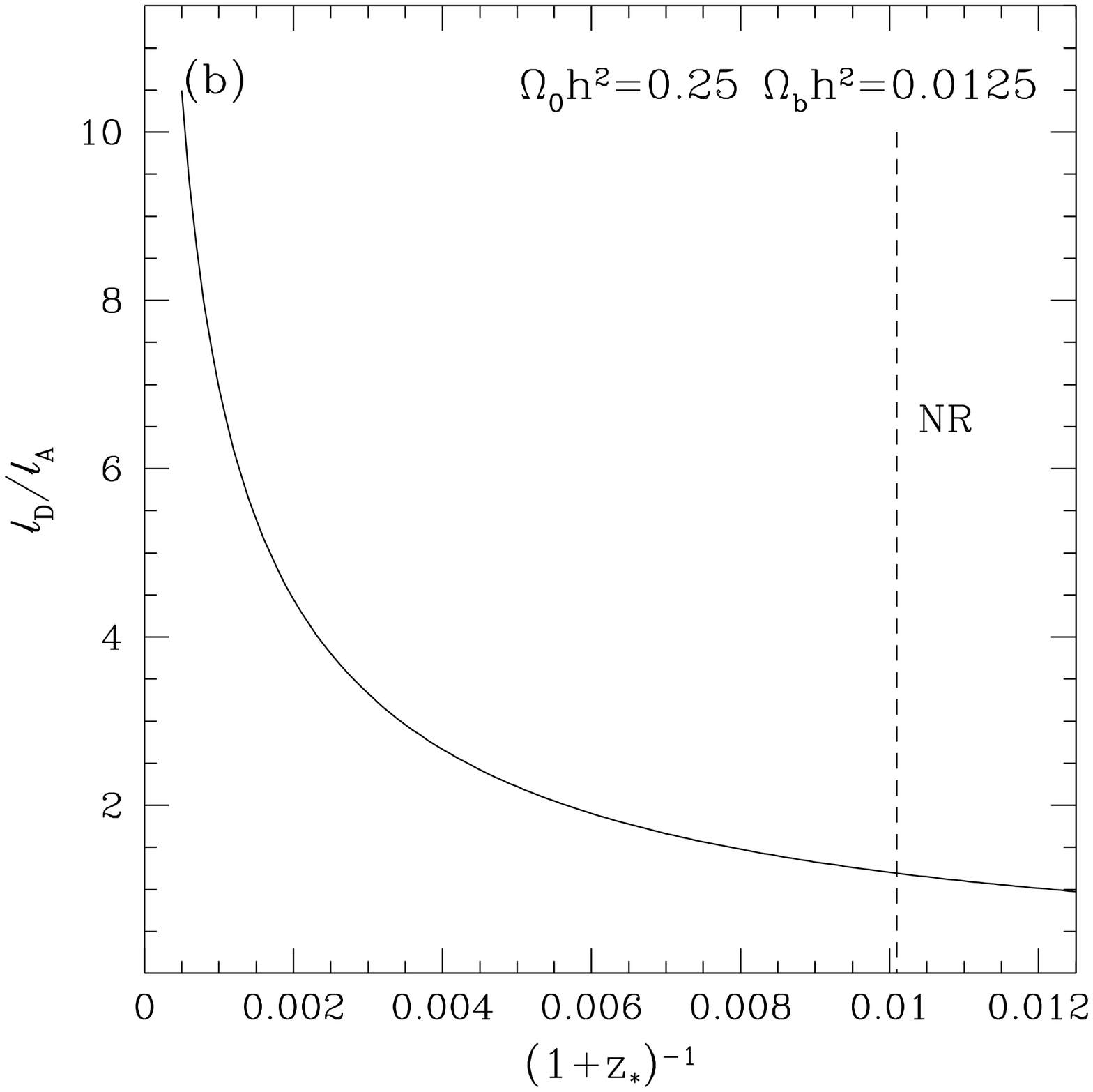}
\end{center}
\vskip 0.75truecm
\caption{\noindent Discriminating exotic cases.  (a) If the baryon content
is far from the BBN value, it will be reflected in the ratio of the damping
scale to peak spacing $\ell_D/\ell_A$, which gives roughly the number of
observable peaks.
(b) Likewise, $\ell_D/\ell_A$ will also detect any delay in recombination.
For simplicity, we have here assumed that the universe recombines
instantaneously at $z_*$.  A realistic recombination scenario will lower
$\ell_D/\ell_A$.
If recombination occurs late, the model becomes identical to the
no-recombination (NR) scenario in its prediction for the CMB.}
\label{fig:exotic}
\end{figure*}

The benefit of this test is that it is entirely independent of the
nature of the fluctuations (see Fig.~\ref{fig:angle}b). 
As long as the baryon fluctuations are linear, the random walk
scale depends only on the background baryon density, ionization 
fraction, and  expansion rate, not on the fluctuations themselves.
The main drawback is that it is difficult to measure accurately.
The signature of diffusion damping is a sharp exponential cutoff in
$\ell$ at the diffusion scale (see Fig.~\ref{fig:angle}a). 
Although this exponential shape is essentially unique, secondary effects
such as gravitational redshifts between last scattering and the present
(ISW effect) can quickly overwhelm the signal making it difficult to measure.
How much these factors will degrade the measurement of the
curvature will vary from model to model.  In inflationary models, both
this effect and various other secondary anisotropies are small enough
that $\ell_D$ should be measurable \cite{HuWhi}.

Another concern is that the damping scale is quite sensitive to
the background baryon content, expansion rate and thermal history.
In Fig.~\ref{fig:omega}b, we show that for reasonable values
of $\Omega_b h^2$ and $h$ this will not prevent us from distinguishing
between $\Omega_0 \approx 0.3$ and $1.0$ models.    

Interestingly the benefits and drawbacks of the peak spacing and damping
tail tests are complementary.  
The peak spacing is easily measured and relatively robust to changes
in the other background parameters, but not fully immune to radical 
behavior in the model for the gravitational fluctuations.  The
damping tail is immune to such effects but is more difficult
to measure and suffers more from uncertainties in the other background
parameters.

Ideally, we would like to measure both quantities. 
By combining these two tests, we have
an important consistency check on the underlying assumptions
of the model and a discriminator against truly exotic models.  
For example, the baryon content $\Omega_b h^2$ could be far
from the BBN value or recombination 
could be delayed by early energy injection from decaying particles
or non-linear structure formation.\footnote{Late
reionization does not interfere with these tests since it mainly
suppresses power uniformly on small scales without changing
the peak spacing or damping tail location.}
Since these exotic possibilities affect the two scales differently,
a useful discriminator is the ratio of the two scales $\ell_D/\ell_A$
(see Fig.~\ref{fig:exotic}).  If this ratio is anomalous, it is 
a clear indication that one of the fundamental assumptions is
invalid.  If one knew from external information which assumption 
that is (e.g.~baryon content, radiation content, thermal history,
etc.) then accurate measurements of the curvature could again
be made.  

The full acoustic signature in the CMB provides additional consistency
checks.  The Compton drag
of the baryons on the photons tends to enhance the fluctuations
inside gravitational potential wells.  For the acoustic oscillations,
this implies that compressional phases will be enhanced over
rarefaction phases leading to an alternating series of relative
peak heights  (see Fig.~\ref{fig:angle}b).  
If the baryon content is far lower than the 
BBN value, then this effect is too weak
to modulate the peak heights.  If it is far higher then it is
so strong that the rarefaction phases will not appear as peaks
in the anisotropy at all \cite{BigPaper}.  This could also
occur if recombination is delayed such that the ratio of baryon
to photon densities is higher at last scattering.
Since the modulation of the peaks is a unique signature of 
baryons at last scattering, its detection would provide an 
important consistency check on the assumptions underlying the
curvature measurement.

In summary, the acoustic signature in the CMB anisotropy spectrum
provides a sufficient number of features such that a curvature
measurement which is essentially robust to the nature of the 
fluctuations, the other background parameters, and thermal history
may be constructed.  The full battery of tests will require complete
information on the acoustic signal -- from the first peak
all the way through to the damping tail.  

\acknowledgements{We would like to thank F. Bouchet and all of the
organizers for a pleasant and productive meeting. 
W.H.~was supported by the NSF and WM Keck Foundation.}

\end{document}